\documentclass[prb,aps,twocolumn,superscriptaddress,floatfix,citeautoscript]{revtex4-2}
\usepackage{graphicx,rotating,subfigure,amsmath,amsfonts,amssymb,delarray,color}
\usepackage{hyperref}
\usepackage{xcolor}

\usepackage{soul}
\usepackage{dsfont}
\usepackage[T1]{fontenc}
\usepackage{physics}
\usepackage{multirow}
\usepackage{comment} 
\usepackage{ulem}

\def\12{\frac{1}{2}}

\begin{document}
\title{Comment on\\ "Resonance-induced growth of number entropy in strongly disordered systems"}

\author{Maximilian Kiefer-Emmanouilidis}
\affiliation{Department of Physics and Research Center OPTIMAS, University Kaiserslautern, 67663 Kaiserslautern, Germany}
\affiliation{Department of Physics and Astronomy, University of Manitoba, Winnipeg R3T 2N2, Canada}
\author{Razmik Unanyan}
\affiliation{Department of Physics and Research Center OPTIMAS, University Kaiserslautern, 67663 Kaiserslautern, Germany}
\author{Michael Fleischhauer}
\affiliation{Department of Physics and Research Center OPTIMAS, University Kaiserslautern, 67663 Kaiserslautern, Germany}
\author{Jesko Sirker}
\affiliation{Department of Physics and Astronomy, University of Manitoba, Winnipeg R3T 2N2, Canada}
\affiliation{Manitoba Quantum Institute, University of Manitoba, Winnipeg R3T 2N2, Canada}

\date{\today}

\begin{abstract}
We comment on the recent paper by Ghosh and Žnidarič  (Phys. Rev. B \textbf{105}, 144203 (2022)) which studies the growth of the number entropy $S_N$ in the Heisenberg model with random magnetic fields after a quantum quench. The authors present arguments for an intermediate power-law growth in time $t$ and a sub-ergodic saturation value, claiming consistency of their results with many-body localization (MBL) for strong disorder. We show that these interpretations are inconsistent with other recent studies and discuss specific issues with the analysis of the numerical data. We point out, in particular, that (i) the saturation values $\widetilde{S}_N(L,W)$ for fixed length $L$ are only bounded from above by 'the ergodic value' and are already far below this value for $W\ll 1$. Furthermore, the saturation values can show non-monotonic scaling with $L$. (ii) Power-law fits $S_N(t)\sim 1/t^\alpha$---with $\alpha=1$ expected based on the resonance model described in the paper---yield a system-size dependent exponent $\alpha$ while fits $S_N\sim \frac{1}{W^3}\ln\ln t$ do hold independent of system size and over several orders of magnitude in time. (iii) We also argue that for the cases where the effective resonance model works best and predicts a saturation of the number entropy, the same applies to the von-Neumann entropy, i.e.~the dynamics at the considered scales is of single particle type and unrelated to MBL.
\end{abstract}

\maketitle

\section{Introduction}

In a recent paper \cite{ghosh2022resonance}, Ghosh and Žnidarič investigate quench dynamics in the Heisenberg model with random magnetic fields. As an indicator for a possible MBL phase, they consider the disorder averaged number entropy $S_N$ obtained from the reduced density matrix after dividing the system in two halves $A$ and $B$ and tracing out one of them. The reduced density matrix then has block structure \footnote{Note that this structure is shown incorrectly in Fig.~1 of \cite{ghosh2022resonance}. The blocks are along the diagonal and there is also a block with $n_A=0$.} with respect to the number of particles $n_A$ in subsystem $A$. The entanglement entropy $S$ can be written as $S=S_N+S_c$ with the number entropy given by $S_N=-\sum_{p_{n_A}} p_{n_A}\ln p_{n_A}$ where $p_{n_A}$ is the probability to find $n_A$ particles in subsystem $A$. $S_c$ is the configurational entropy. We agree with the authors that $S_N$ is a useful indicator for a possible MBL transition: While $S_N(t)$ will grow without bounds after a quantum quench in the thermodynamic limit (TDL) if the system is not localized, an MBL phase is characterized by $S_N(t\to\infty)=\mbox{const}$ while $S(t)\sim\ln t$.  

Before we discuss specific data and conclusions in \cite{ghosh2022resonance}, we want to summarize some important general points: (i) In a finite system, both $S$ and $S_N$ will saturate. A finite size scaling is therefore crucial. It is also crucial to study {\it both} $S$ and $S_N$. While one can find initial states where $S_N$ seems to saturate quickly for the system sizes and times accessible to exact diagonalizations (ED), this by itself does not point to MBL. One also needs to show that in this regime $S\sim\ln t$. If $S$ grows slower than $\ln t$ or also saturates quickly then this merely points to insufficient system sizes and a possible dominance of single particle physics on these time and length scales. (ii) For disorder strengths $W\gtrsim 10$, $S_N(t)$ is difficult to analyze by ED because the time $t_d$, where the results start to deviate from the TDL, scales exponentially both with system size $L$ and disorder strength $W$ for typical initial states, see Ref.~\cite{KieferUnanyan4}. Double precision though is insufficient to study dynamics for $t\gtrsim 10^{14}$. Furthermore, we have found in the same paper that $S_N\sim\frac{1}{W^3}\ln\ln t$. I.e., the prefactor and thus the growth of $S_N(t)$ becomes very small for large $W$ and needs to be analyzed carefully. (iii) For the saturation value $\tilde S_N(L,W)$ we note first that for any finite $L$, $\tilde S_N$ will be an analytic function of $W$. In particular, it will continuously decrease as a function of $W$ and will be below 'the ergodic value' $S_N\sim\frac{1}{2}\ln(e\pi L/8)$ even for very small disorder. For a generic interacting system, the saturation value of the number entropy averaged over all initial half-filled product  states is bounded from above by the entropy of the hypergeometric distribution
which in the asymptotic limit of large $L$ approaches $\frac{1}{2}\ln\left(  e\pi L/8\right)$. 
Second, $\tilde S_N(L,W)$ {\it does not} have to be a monotonically increasing function of $L$. The scaling does depend on the initial state. In particular, if an average over all initial product states is considered then 
already in the non-interacting case there is a well-understood $\sim 1/L$ correction (see Appendix A and Fig.~11 in \cite{KieferUnanyan5}) which competes with any potential increase in the particle number fluctuations due to interactions. In cases where such a $1/L$ correction is absent, for example for the initial N\'eel state \cite{KieferUnanyan5}, $\tilde S_N(L,W)\sim\ln L$ for fixed $W$. (iv) Initial states such as the domain wall state or partial domain walls ('q-states' in \cite{ghosh2022resonance}) are problematic starting points for a search of an MBL phase: in small finite systems only a small number of particles at the edge of the domain wall are initially able to move. Especially for large $W$, the dynamics in small systems is then dominated by single-particle physics. (v) In Refs.~\cite{KieferUnanyan1,KieferUnanyan2,KieferUnanyan3,KieferUnanyan4,KieferUnanyan5} we have established through finite-size scaling that $S_N(t)\sim \frac{1}{W^3}\ln\ln t$ for $W\lesssim 10$, a scaling which holds over 4 orders of magnitude in time for the largest system sizes studied. We have also established that in the same regime particle fluctuations in the subsystem of more than one particle are present as can be seen, for example, by studying the Hartley number entropy, see \cite{KieferUnanyan4}. This points to the absence of localization in this regime. These results are consistent with the most recent studies \cite{MorningstarHusev2,SelsMarkovian2021} which find no localization for $W\lesssim 20$ by using completely independent indicators. Note that the latter study is based on the extension of the method proposed in the former to larger system sizes. These results are not adequately described in \cite{ghosh2022resonance} and most of the data presented in \cite{ghosh2022resonance} fall into the regime which is not localized according to these studies. 

Let us now concentrate on some of the most relevant issues. In Sec.~\ref{Saturation} we show that the saturation value $\tilde S_N(L,W)$ is monotonically decreasing with $W$ and always smaller than the 'ergodic value' even for very small disorder. We show that the same is also true in the off-diagonal disorder model which is known to be not localized. In Sec.~\ref{PowerLaw} we present data for a quench from the N\'eel state which show that in a power-law fit $S_N(t)=c_1-c_2/t^\alpha$ for intermediate times, with $c_{1,2}$ constants which depend on $L$, the exponent $\alpha$ is not close to $1$ as predicted in \cite{ghosh2022resonance} but rather also depends on length with $\alpha$ monotonically decreasing with increasing $L$. In contrast, $S_N=c_1+c_2\ln\ln t$ describes the data at intermediate times for all lengths with $c_{1,2}$ independent of $L$. In Sec.~\ref{States} we then discuss the choice of initial states and argue, in particular, that for small system sizes quenches from the domain wall state or large q-states are more akin to local quenches. The $|\mathcal{I}\rangle$ state, on the other hand, leads to dynamics where the time range with $S\sim\ln t$ is extremely small for system sizes accessible by ED. In Sec.~\ref{Resonances} we demonstrate that for the initial states and large disorder strengths where the authors find that a simple resonance model describes $S_N(t)$, the same holds also true for the entropy $S(t)$ and $S\approx S_N$. The dynamics stays entirely local in these cases and can be explained by single particle physics. 
\section{Saturation value of number entropy}
\label{Saturation}
In \cite{ghosh2022resonance} the authors write "Saturation
values of $S_N$ are in all cases small and far from being
ergodic. For example, a random half-filled state to which
one would converge at long times in an ergodic system has $S_N\approx\frac{1}{2}\ln (e\pi L/8)$, $\cdots$". This is shown in Fig.~4 of their paper and argued to confirm the lack of ergodicity.

\begin{figure}[h!]
    \centering
    \includegraphics[width=0.9\columnwidth]{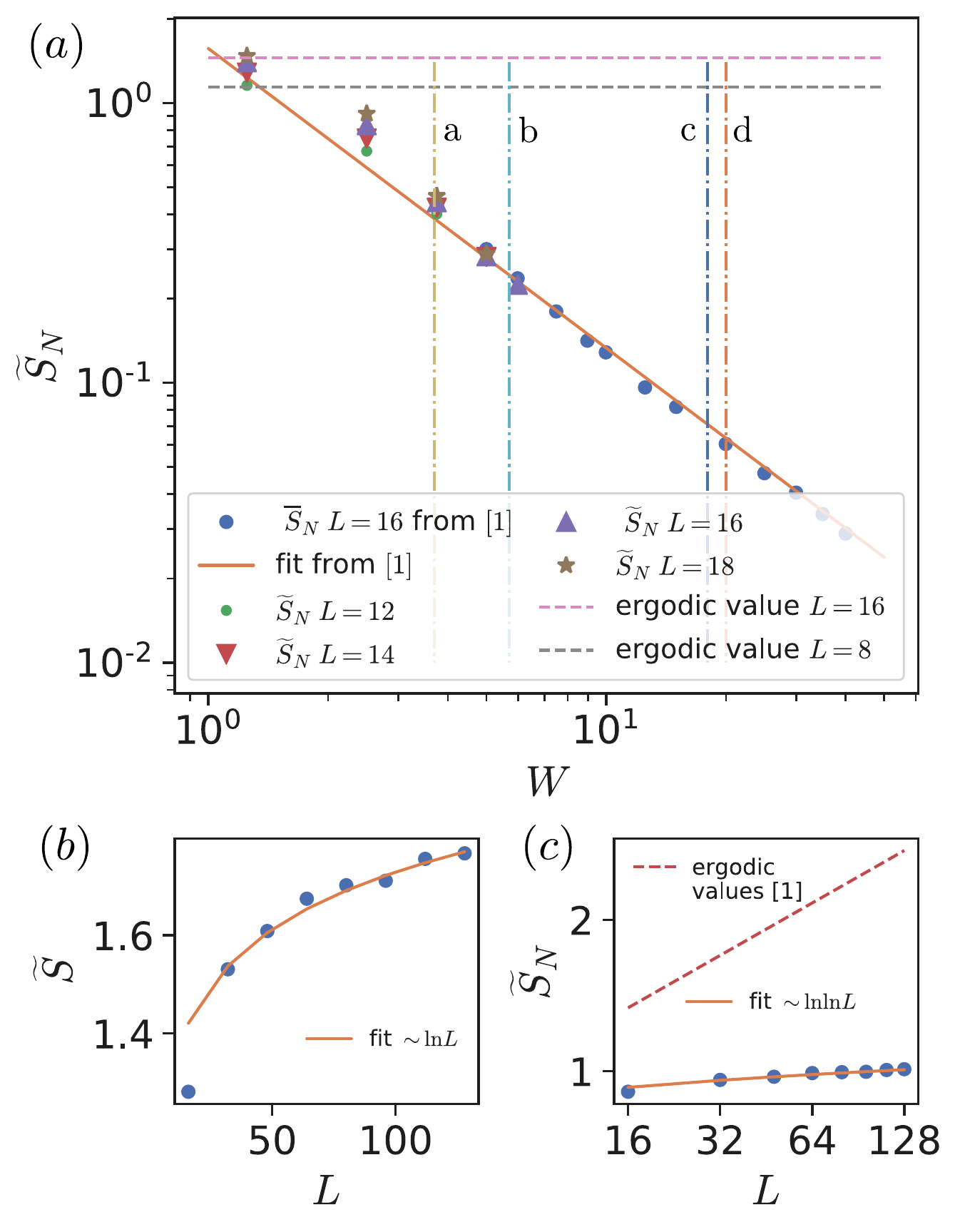}
    \caption{(a) Saturation values $\overline{ S}_N$ as function of disorder strength taken from \cite{ghosh2022resonance} and supplemented with values $\widetilde{ S}_N$ for smaller $W$ taken from \cite{KieferUnanyan2}. Note that $\overline{ S}_N$ and $\widetilde{ S}_N$ become indistinguishable at large $W$ on this scale.
    The vertical lines indicate different theoretical predictions of the
    critical disorder strength. (a - Ref.\cite{Luitz2015}, b - Ref.\cite{MorningstarHuse}, c - Ref.\cite{MorningstarHusev2}, d - Ref. \cite{SelsMarkovian2021}). (b) $\widetilde{S}(L)\equiv S(t\rightarrow \infty)$ for the off-diagonal disorder case with a fit $\widetilde{S}\sim \ln L$ \cite{IgloiSzatmari,VoskAltman,RefaelPekker,Zhao2016},
    and (c) the corresponding $\widetilde{S}_N(L)\sim\ln\ln L$ \cite{KieferUnanyan1}.  For our data shown in (a) we have averaged over 10000 disorder realizations and initial states for $L\leq 14$ and 3000 for $L>14$. The data presented in (b,c) are averaged over at least 10000 disorder realizations and initial states.
    }
    \label{Fig_sat}
\end{figure}

In Fig.\ref{Fig_sat}(a) we compare the saturation values of the number entropy taken from \cite{ghosh2022resonance}---with values for smaller disorder added---with the 'ergodic value' above, which is, in fact, the value expected for a hypergeometric distribution of particles. 
In \cite{ghosh2022resonance} $\widetilde S_N$ is the long-time average of $S_N$ which has been calculated by averaging data between $t\sim10^6-10^7$ and $\overline{S}_N=-\sum_{n_A=0} \overline{p}_{n_A} \mathrm{ln} \overline{p}_{n_A}$, where $\overline{p}_{n_A}$ are the long time averages of $p_{n_A}$, and $\widetilde S_N\leq\overline S_N$ holds.
One notices that the saturation values $\tilde S_N$ are bound from above by the hypergeometric one including in the regime of very small disorder where no MBL is expected. Secondly, the decrease of the saturation value with disorder strength follows a smooth power-law with no indication of a phase transition. We have indicated the predicted transition points from several publications in the TDL. We therefore believe that Fig.\ref{Fig_sat}(a) rather supports the interpretation of a smooth crossover and that being below the hypergeometric value cannot be used to argue in favour of localization. This criticism is supported by Fig.~\ref{Fig_sat}(c) which shows that in the off-diagonal disorder case, known to be not localized, $\tilde S_N$ is also well below the hypergeometric value for all finite system sizes. 

Let us finally note that a decrease of the saturation value with system size if one averages over all initial product states as shown in Fig.~4(b) in \cite{ghosh2022resonance} has been observed previously \cite{BarLevLuitz,KieferUnanyan3}. It is explained by a $1/L$ correction due to states with large number fluctuations which is already present in the non-interacting case \cite{KieferUnanyan5}. To study the effect of interactions, $S_N$ relative to the non-interacting case needs to be considered in this case. 

\section{Power-law time-evolution of the number entropy}
\label{PowerLaw}
The main result of Ref.~\cite{ghosh2022resonance} appears to be the prediction that the number entropy increases like a power law at intermediate times, followed by a system-size independent saturation. More precisely, the authors write in Eq.~(14) of \cite{ghosh2022resonance} that
\begin{equation}
\label{WrongTheory}
S_N(t) = \mbox{const} - B/t    
\end{equation}
with a constant $B$. Using a quench from the N\'eel state as an example, we show that this prediction is false. Instead, the growth is well described for up to 4 orders of magnitude in time by 
\begin{equation}
\label{CorrectTheory}
S_N(t) = c_1+c_2\ln\ln t   
\end{equation}
with constants $c_{1,2}$ which do not depend on system size. I.e., the growth is consistent with our earlier results presented in \cite{KieferUnanyan2,KieferUnanyan3, KieferUnanyan4,KieferUnanyan5}. 

\begin{figure}[h!]
    \centering
    \includegraphics[width=0.9\columnwidth]{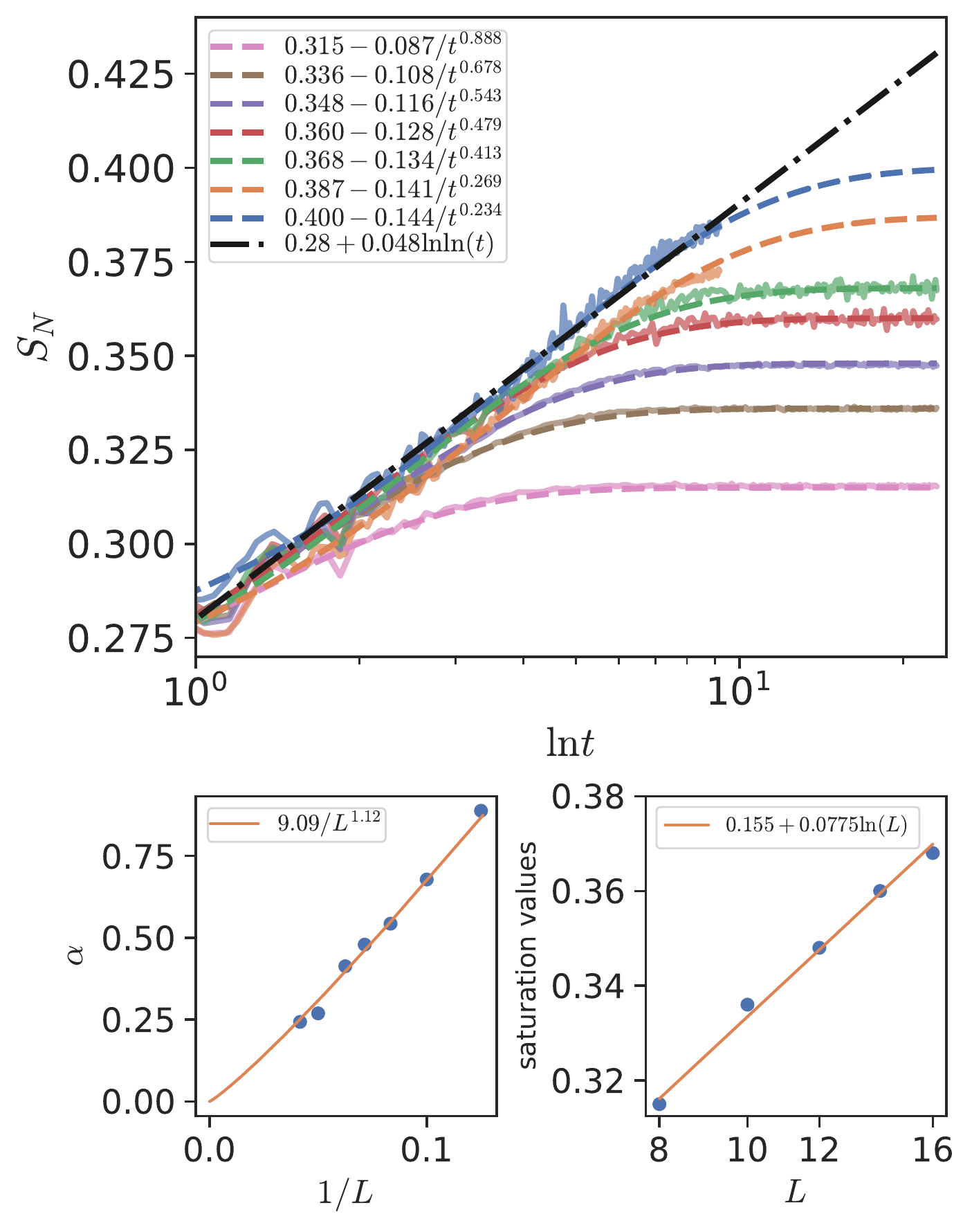}
    \caption{Quench from the N\'eel state for $W=5$ and system sizes $L=8,10,\cdots 16, 20, 24$. The left lower panel shows the exponent $\alpha$ of power-law fits, the right lower panel the saturation value as a function of length. We have computed $2\times 10^5$ disorder realizations for $L=8$ and $L=10$, 80000 for $L=12$, 4000 for $L=14$, 3000 for $L=16$, 5300 for $L=20$, and 1406 for $L=24$. For $L>16$ we used a Trotter-Suzuki decomposition as described in \cite{KieferUnanyan2} with $\delta t = 0.035$.}
    \label{Fig1}
\end{figure}

In Fig.~\ref{Fig1} we show the number entropy for a quench from the N\'eel state and a disorder strength $W=5$. 
We note that reliable fits of the time dependence of $S_N(t)$ become virtually impossible for $W\gtrsim 10$ because the prefactor in Eq.~\eqref{CorrectTheory} scales as $c_2\sim 1/W^3$ and the saturation time increases exponentially with $W$, see Ref.~\cite{KieferUnanyan4} for details. Therefore double precision calculations are no longer sufficient for $W\gtrsim 10$ and if multi-precision is used, a huge number of samples would be needed to resolve the small increase. While fits by a power law do work reasonably well close to the finite-size saturation values, the power-law exponent $\alpha$ does depend on system size $L$ and appears to approach zero for $L\to\infty$. Furthermore, the saturation value is not constant but rather increases $\sim \ln L$ (see also \cite{KieferUnanyan5}). On the other hand, all data fall {\it onto a single} $\ln\ln t$ curve before saturation due to the finite size of the systems sets in.
\begin{figure}[ht!]
    \centering
    \includegraphics[width=0.9\columnwidth]{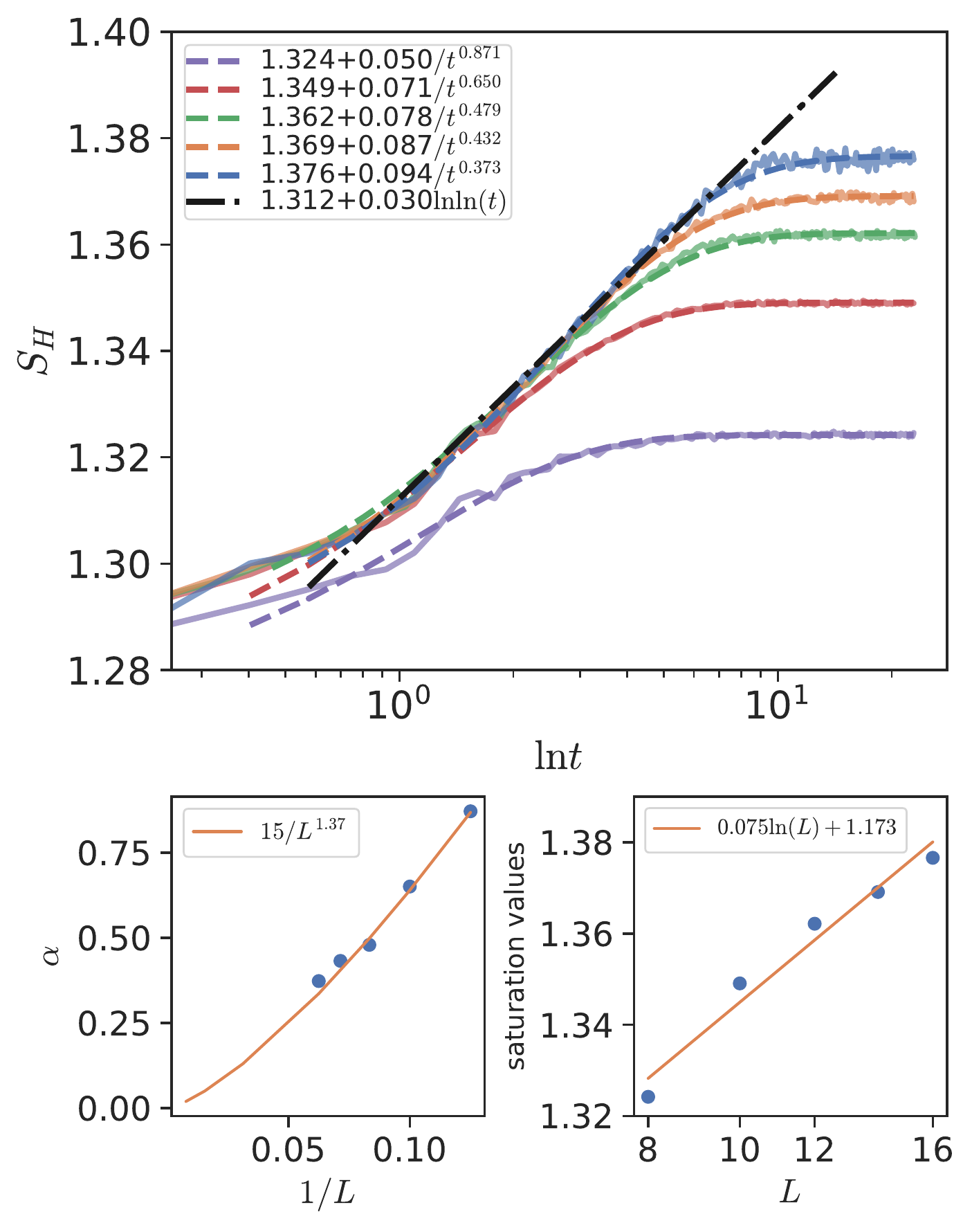}
    \caption{Same as in Fig.~\ref{Fig1} but for the Hartley entropy $S_H$ and for larger disorder $W=8$. We have computed $ 10^5$ disorder realizations for $L=8$ and $L=10$, 20000 for $L=12$, 10000 for $L=14$, 2000 for $L=16$.}
    \label{Fig2}
\end{figure}

In addition, we also show data for the Hartley entropy and $W=8$ in Fig.~\ref{Fig2}. The Hartley entropy is defined in Ref.~\cite{KieferUnanyan3} and more easily allows to extract the scaling behaviour in time for very strongly disordered systems. The simple reason is that the Hartley entropy is larger than the number entropy so a proper scaling can already be extracted using a much smaller amount of samples than what would be needed for the number entropy. As for the number entropy, we find that the effective power-law exponent $\alpha$ is not equal to one as predicted by the theory in Ref.~\cite{ghosh2022resonance} but rather tends to zero with increasing system size. The saturation values of $S_H$ are also not constant but again increase logarithmically with system size. To summarize, the theory \eqref{WrongTheory} does not describe the data. They are instead well described by \eqref{CorrectTheory}.

\section{Choice of initial state}
\label{States}
It is important to stress that Anderson localization is not about the localization of a particle whose motion is already restricted to a finite region of space. Rather, the particle is in principle able to move infinitely far away from its initial position; what is stopping it from doing so is that the effective coupling to another site falls off exponentially with distance while the energy mismatch only falls off like one over distance in one dimension \cite{50yearsAnderson}. Similarly, to study the possible existence of MBL one has to understand the occurrence of many-body resonances and avalanche instabilities which are global, not local, properties of the system \cite{MorningstarHuse}.

To numerically investigate the question whether or not an MBL phase in the Heisenberg model
does exist, it is thus important to start from an initial state in which a macroscopic number of particles are able to move. I.e., initial states which do have {\it extensive} energy fluctuations $\Delta E^2$. This is the case, for example, for the N\'eel state used above and also for the vast majority of other product states. For this reason, almost all numerical studies of the MBL problem so far use the N\'eel state and/or randomly drawn product states as initial states. The authors of Ref.~\cite{ghosh2022resonance}, on the other hand, choose in their numerical examples to support the resonance model in Sec.~III very special product states as, for example, the domain wall state which has $O(1)$ energy fluctuations: only the particle at the edge of the domain wall is initially able to move. 
In particular, for all product states which are eigenstates of the total particle number operator $\hat N$ the energy fluctuations can be simply quantified by the number of kinks in the state: $\Delta E^2\sim N_\mathrm{kinks}$. The Neel state has $N_\mathrm{kinks}=L-1$ and thus has extensive energy fluctuations while the domain wall state has $N_\mathrm{kinks}=1$ and therefore intensive energy fluctuations.

Another way to look at this issue is in terms of the difference between a global and a local quench: the domain wall state is an eigenstate of the Heisenberg model with the bond connecting the two domains removed. So the quench is effectively a local one. Typical initial product states, on the other hand, are not eigenstates of the Hamiltonian with a finite number of local modifications. It should be noted, furthermore, that the domain wall state remains partially frozen in a quench using the XXZ model with $\Delta>1$ even without any disorder \cite{Mossel_2010}.  

The q-states considered ---at least for $q> 2$---fall, for the considered small system sizes, also into the category of atypical initial states with small energy fluctuations and an initial dynamics restricted to the boundaries of the finite domain wall. While these states will eventually show extensive fluctuations for fixed q and large system sizes $L$, a finite-size scaling was not attempted in \cite{ghosh2022resonance} and the accessible system sizes are likely too small for the larger q-states to ever see typical dynamics.

The authors also consider the $|\mathcal{I}\rangle$ state which is the tensor product of the independent superposition of all states in subsystems $A$ and $B$ with half the particles in each subsystem. Here we note that while $S_N$ indeed shows only very slow growth, the time range in which $S\sim\ln t$ is also extremely small, see Figs.~13, 14 in \cite{ghosh2022resonance}. Thus these data, in our view, simply show that for this initial state the system sizes accessible in ED are not sufficient to make any statements about the scaling in the TDL.

\section{Resonances and quenches from a domain wall state}
\label{Resonances}

We have argued above that the domain wall state and the q-states are not suitable to investigate MBL physics because they show only local dynamics at strong disorder and for small system sizes. Here we will provide numerical data to support this point, taking the domain wall state as an example.

\begin{figure}[h!]
    \centering
    \includegraphics[width=0.49\columnwidth]{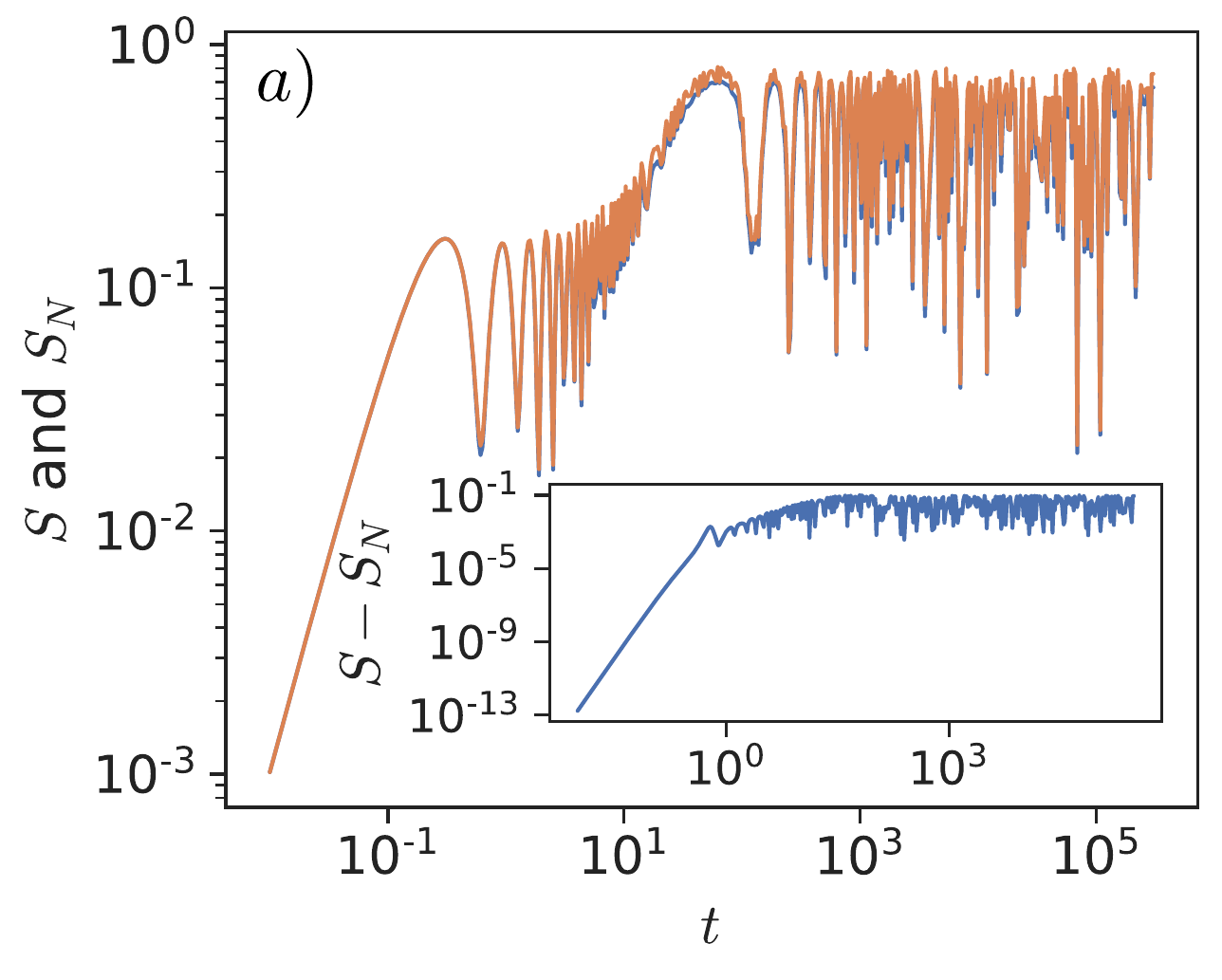}
    \includegraphics[width=0.49\columnwidth]{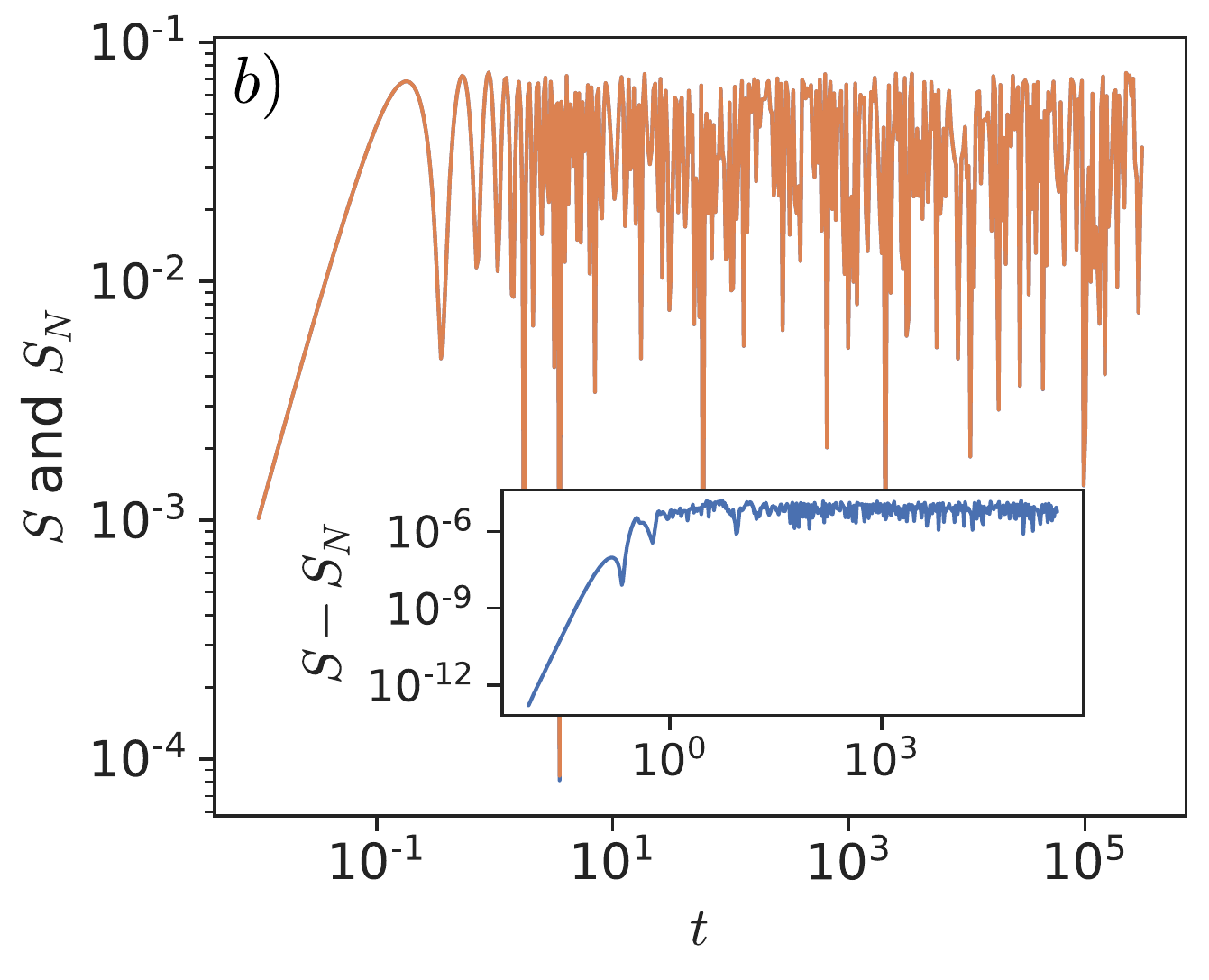}
    \caption{The analogue of Fig.~3 in \cite{ghosh2022resonance}. For a quench from the domain wall state and strong disorder $W=15$, not only the number entropy saturates quickly but also the von-Neumann entropy. (a) Rare disorder realization: A plateau is visible before the saturation value is reached. The inset shows $S-S_N$. (b) Typical disorder realization: The configurational entropy is extremely small. Note that on this scale $S$ and $S_N$ cannot be distinguished, see inset for $S-S_N$. Here typical means that we determined the median of 1000 realizations and selected one realization which oscillates around the median value. }
    \label{Fig_rare}
\end{figure}

The authors of Ref.~\cite{ghosh2022resonance} base their theory of the growth of $S_N$ on resonances between product states connected by the hopping of a single particle. As an example, they show in Fig.~3 of their paper results for quenches from the domain wall state for strong disorder for different specific disorder realizations. Depending on the chosen realization, 
jumps of $S_N(t)$ occur at different times that can be associated to certain resonances. The authors argue that averaging over disorder configurations then smooth out the jumps leading to an intermediate slow power-law increase of the number entropy.
They argue that in this case, only very few states are involved in the dynamics of the system. We agree with this statement, however, it holds true also for the von Neumann entropy $S$, which unfortunately was not shown in \cite{ghosh2022resonance}. Both quantities are nearly indistinguishable for the parameters chosen and saturate at short times. This is particularly obvious if one considers the configurational entropy $S_C = S-S_N$, see Fig.\ref{Fig_rare}. 
$S_c$ is extremely small for typical disorder configurations. We conclude that for this initial state, large disorder values $W=15$ and higher, and the considered system sizes and time scales, only single particle dynamics is visible. The system is essentially Anderson localized and the entanglement entropy is caused by number fluctuations alone. While the oscillations at long time can indeed be studied using an effective $m$-state model, the observed behavior is not related to many-body localization.

\section*{Acknowledgement}
M.K-E., R.U., and M.F. acknowledge financial support from the Deutsche Forschungsgemeinschaft (DFG) via SFB TR 185, Project No.277625399. J.S. acknowledges support by the National Science and Engineering Council (NSERC, Canada) and by the DFG via Research Unit FOR 2316. The numerical simulations were executed on the GPU nodes of the high performance cluster “Elwetritsch” at the University of Kaiserslautern which is part of the “Alliance of High Performance Computing Rheinland-Pfalz” (AHRP) and on Compute Canada high-performance clusters. We kindly acknowledge the support of the RHRK and of Compute Canada.

\begin{thebibliography}{18}%
\makeatletter
\providecommand \@ifxundefined [1]{%
 \@ifx{#1\undefined}
}%
\providecommand \@ifnum [1]{%
 \ifnum #1\expandafter \@firstoftwo
 \else \expandafter \@secondoftwo
 \fi
}%
\providecommand \@ifx [1]{%
 \ifx #1\expandafter \@firstoftwo
 \else \expandafter \@secondoftwo
 \fi
}%
\providecommand \natexlab [1]{#1}%
\providecommand \enquote  [1]{``#1''}%
\providecommand \bibnamefont  [1]{#1}%
\providecommand \bibfnamefont [1]{#1}%
\providecommand \citenamefont [1]{#1}%
\providecommand \href@noop [0]{\@secondoftwo}%
\providecommand \href [0]{\begingroup \@sanitize@url \@href}%
\providecommand \@href[1]{\@@startlink{#1}\@@href}%
\providecommand \@@href[1]{\endgroup#1\@@endlink}%
\providecommand \@sanitize@url [0]{\catcode `\\12\catcode `\$12\catcode
  `\&12\catcode `\#12\catcode `\^12\catcode `\_12\catcode `\%12\relax}%
\providecommand \@@startlink[1]{}%
\providecommand \@@endlink[0]{}%
\providecommand \url  [0]{\begingroup\@sanitize@url \@url }%
\providecommand \@url [1]{\endgroup\@href {#1}{\urlprefix }}%
\providecommand \urlprefix  [0]{URL }%
\providecommand \Eprint [0]{\href }%
\providecommand \doibase [0]{https://doi.org/}%
\providecommand \selectlanguage [0]{\@gobble}%
\providecommand \bibinfo  [0]{\@secondoftwo}%
\providecommand \bibfield  [0]{\@secondoftwo}%
\providecommand \translation [1]{[#1]}%
\providecommand \BibitemOpen [0]{}%
\providecommand \bibitemStop [0]{}%
\providecommand \bibitemNoStop [0]{.\EOS\space}%
\providecommand \EOS [0]{\spacefactor3000\relax}%
\providecommand \BibitemShut  [1]{\csname bibitem#1\endcsname}%
\let\auto@bib@innerbib\@empty
\bibitem [{\citenamefont {Ghosh}\ and\ \citenamefont
  {{\v{Z}}nidari{\v{c}}}(2022)}]{ghosh2022resonance}%
  \BibitemOpen
  \bibfield  {author} {\bibinfo {author} {\bibfnamefont {R.}~\bibnamefont
  {Ghosh}}\ and\ \bibinfo {author} {\bibfnamefont {M.}~\bibnamefont
  {{\v{Z}}nidari{\v{c}}}},\ }\bibfield  {title} {\bibinfo {title}
  {Resonance-induced growth of number entropy in strongly disordered systems},\
  }\href
{https://doi.org/10.1103/PhysRevB.105.144203} {\bibfield  {journal} {\bibinfo  {journal} {Physical Review
  B}\ }\textbf {\bibinfo {volume} {105}},\ \bibinfo {pages} {144203} (\bibinfo
  {year} {2022})}\BibitemShut {NoStop}%
\bibitem [{Note1()}]{Note1}%
  \BibitemOpen
  \bibinfo {note} {Note that this structure is shown incorrectly in Fig.~1 of
  \cite {ghosh2022resonance}. The blocks are along the diagonal and there is
  also a block with $n_A=0$.}\BibitemShut {Stop}%
\bibitem [{\citenamefont {Kiefer-Emmanouilidis}\ \emph
  {et~al.}(2021{\natexlab{a}})\citenamefont {Kiefer-Emmanouilidis},
  \citenamefont {Unanyan}, \citenamefont {Fleischhauer},\ and\ \citenamefont
  {Sirker}}]{KieferUnanyan4}%
  \BibitemOpen
  \bibfield  {author} {\bibinfo {author} {\bibfnamefont {M.}~\bibnamefont
  {Kiefer-Emmanouilidis}}, \bibinfo {author} {\bibfnamefont {R.}~\bibnamefont
  {Unanyan}}, \bibinfo {author} {\bibfnamefont {M.}~\bibnamefont
  {Fleischhauer}},\ and\ \bibinfo {author} {\bibfnamefont {J.}~\bibnamefont
  {Sirker}},\ }\bibfield  {title} {\bibinfo {title} {Unlimited growth of
  particle fluctuations in many-body localized phases},\ }\href
  {https://doi.org/10.1016/j.aop.2021.168481} {\bibfield  {journal} {\bibinfo
  {journal} {Annals of Physics}\ ,\ \bibinfo {pages} {168481}} (\bibinfo {year}
  {2021}{\natexlab{a}})}\BibitemShut {NoStop}%
\bibitem [{\citenamefont {Kiefer-Emmanouilidis}\ \emph
  {et~al.}(2022)\citenamefont {Kiefer-Emmanouilidis}, \citenamefont {Unanyan},
  \citenamefont {Fleischhauer},\ and\ \citenamefont {Sirker}}]{KieferUnanyan5}%
  \BibitemOpen
  \bibfield  {author} {\bibinfo {author} {\bibfnamefont {M.}~\bibnamefont
  {Kiefer-Emmanouilidis}}, \bibinfo {author} {\bibfnamefont {R.}~\bibnamefont
  {Unanyan}}, \bibinfo {author} {\bibfnamefont {M.}~\bibnamefont
  {Fleischhauer}},\ and\ \bibinfo {author} {\bibfnamefont {J.}~\bibnamefont
  {Sirker}},\ }\bibfield  {title} {\bibinfo {title} {{Particle fluctuations and
  the failure of simple effective models for many-body localized phases}},\
  }\href {https://doi.org/10.21468/SciPostPhys.12.1.034} {\bibfield  {journal}
  {\bibinfo  {journal} {SciPost Phys.}\ }\textbf {\bibinfo {volume} {12}},\
  \bibinfo {pages} {34} (\bibinfo {year} {2022})}\BibitemShut {NoStop}%
\bibitem [{\citenamefont {Kiefer-Emmanouilidis}\ \emph
  {et~al.}(2020{\natexlab{a}})\citenamefont {Kiefer-Emmanouilidis},
  \citenamefont {Unanyan}, \citenamefont {Sirker},\ and\ \citenamefont
  {Fleischhauer}}]{KieferUnanyan1}%
  \BibitemOpen
  \bibfield  {author} {\bibinfo {author} {\bibfnamefont {M.}~\bibnamefont
  {Kiefer-Emmanouilidis}}, \bibinfo {author} {\bibfnamefont {R.}~\bibnamefont
  {Unanyan}}, \bibinfo {author} {\bibfnamefont {J.}~\bibnamefont {Sirker}},\
  and\ \bibinfo {author} {\bibfnamefont {M.}~\bibnamefont {Fleischhauer}},\
  }\bibfield  {title} {\bibinfo {title} {{Bounds on the entanglement entropy by
  the number entropy in non-interacting fermionic systems}},\ }\href
  {https://doi.org/10.21468/SciPostPhys.8.6.083} {\bibfield  {journal}
  {\bibinfo  {journal} {SciPost Phys.}\ }\textbf {\bibinfo {volume} {8}},\
  \bibinfo {pages} {83} (\bibinfo {year} {2020}{\natexlab{a}})}\BibitemShut
  {NoStop}%
\bibitem [{\citenamefont {Kiefer-Emmanouilidis}\ \emph
  {et~al.}(2020{\natexlab{b}})\citenamefont {Kiefer-Emmanouilidis},
  \citenamefont {Unanyan}, \citenamefont {Fleischhauer},\ and\ \citenamefont
  {Sirker}}]{KieferUnanyan2}%
  \BibitemOpen
  \bibfield  {author} {\bibinfo {author} {\bibfnamefont {M.}~\bibnamefont
  {Kiefer-Emmanouilidis}}, \bibinfo {author} {\bibfnamefont {R.}~\bibnamefont
  {Unanyan}}, \bibinfo {author} {\bibfnamefont {M.}~\bibnamefont
  {Fleischhauer}},\ and\ \bibinfo {author} {\bibfnamefont {J.}~\bibnamefont
  {Sirker}},\ }\bibfield  {title} {\bibinfo {title} {Evidence for unbounded
  growth of the number entropy in many-body localized phases},\ }\href
  {https://doi.org/10.1103/PhysRevLett.124.243601} {\bibfield  {journal}
  {\bibinfo  {journal} {Phys. Rev. Lett.}\ }\textbf {\bibinfo {volume} {124}},\
  \bibinfo {pages} {243601} (\bibinfo {year} {2020}{\natexlab{b}})}\BibitemShut
  {NoStop}%
\bibitem [{\citenamefont {Kiefer-Emmanouilidis}\ \emph
  {et~al.}(2021{\natexlab{b}})\citenamefont {Kiefer-Emmanouilidis},
  \citenamefont {Unanyan}, \citenamefont {Fleischhauer},\ and\ \citenamefont
  {Sirker}}]{KieferUnanyan3}%
  \BibitemOpen
  \bibfield  {author} {\bibinfo {author} {\bibfnamefont {M.}~\bibnamefont
  {Kiefer-Emmanouilidis}}, \bibinfo {author} {\bibfnamefont {R.}~\bibnamefont
  {Unanyan}}, \bibinfo {author} {\bibfnamefont {M.}~\bibnamefont
  {Fleischhauer}},\ and\ \bibinfo {author} {\bibfnamefont {J.}~\bibnamefont
  {Sirker}},\ }\bibfield  {title} {\bibinfo {title} {Slow delocalization of
  particles in many-body localized phases},\ }\href
  {https://doi.org/10.1103/PhysRevB.103.024203} {\bibfield  {journal} {\bibinfo
   {journal} {Phys. Rev. B}\ }\textbf {\bibinfo {volume} {103}},\ \bibinfo
  {pages} {024203} (\bibinfo {year} {2021}{\natexlab{b}})}\BibitemShut
  {NoStop}%
\bibitem [{\citenamefont {Morningstar}\ \emph
  {et~al.}(2021{\natexlab{a}})\citenamefont {Morningstar}, \citenamefont
  {Colmenarez}, \citenamefont {Khemani}, \citenamefont {Luitz},\ and\
  \citenamefont {Huse}}]{MorningstarHusev2}%
  \BibitemOpen
  \bibfield  {author} {\bibinfo {author} {\bibfnamefont {A.}~\bibnamefont
  {Morningstar}}, \bibinfo {author} {\bibfnamefont {L.}~\bibnamefont
  {Colmenarez}}, \bibinfo {author} {\bibfnamefont {V.}~\bibnamefont {Khemani}},
  \bibinfo {author} {\bibfnamefont {D.~J.}\ \bibnamefont {Luitz}},\ and\
  \bibinfo {author} {\bibfnamefont {D.~A.}\ \bibnamefont {Huse}},\ }\bibfield
  {title} {\bibinfo {title} {Avalanches and many-body resonances in many-body
  localized systems},\ }\href@noop {} {\bibfield  {journal} {\bibinfo
  {journal} {arXiv:2107.05642v2}\ } (\bibinfo {year} {2021}{\natexlab{a}})},\
  \Eprint {https://arxiv.org/abs/2107.05642} {arXiv:2107.05642
  [cond-mat.dis-nn]} \BibitemShut {NoStop}%
\bibitem [{\citenamefont {Sels}(2021)}]{SelsMarkovian2021}%
  \BibitemOpen
  \bibfield  {author} {\bibinfo {author} {\bibfnamefont {D.}~\bibnamefont
  {Sels}},\ }\bibfield  {title} {\bibinfo {title} {Markovian baths and quantum
  avalanches},\ }\bibfield  {journal} {\bibinfo  {journal}
  {arXiv:2108.10796v1}\ }\href {https://doi.org/10.48550/ARXIV.2108.10796}
  {10.48550/ARXIV.2108.10796} (\bibinfo {year} {2021})\BibitemShut {NoStop}%
\bibitem [{\citenamefont {Luitz}\ \emph {et~al.}(2015)\citenamefont {Luitz},
  \citenamefont {Laflorencie},\ and\ \citenamefont {Alet}}]{Luitz2015}%
  \BibitemOpen
  \bibfield  {author} {\bibinfo {author} {\bibfnamefont {D.~J.}\ \bibnamefont
  {Luitz}}, \bibinfo {author} {\bibfnamefont {N.}~\bibnamefont {Laflorencie}},\
  and\ \bibinfo {author} {\bibfnamefont {F.}~\bibnamefont {Alet}},\ }\bibfield
  {title} {\bibinfo {title} {Many-body localization edge in the random-field
  heisenberg chain},\ }\href {https://doi.org/10.1103/PhysRevB.91.081103}
  {\bibfield  {journal} {\bibinfo  {journal} {Phys. Rev. B}\ }\textbf {\bibinfo
  {volume} {91}},\ \bibinfo {pages} {081103} (\bibinfo {year}
  {2015})}\BibitemShut {NoStop}%
\bibitem [{\citenamefont {Morningstar}\ \emph
  {et~al.}(2021{\natexlab{b}})\citenamefont {Morningstar}, \citenamefont
  {Colmenarez}, \citenamefont {Khemani}, \citenamefont {Luitz},\ and\
  \citenamefont {Huse}}]{MorningstarHuse}%
  \BibitemOpen
  \bibfield  {author} {\bibinfo {author} {\bibfnamefont {A.}~\bibnamefont
  {Morningstar}}, \bibinfo {author} {\bibfnamefont {L.}~\bibnamefont
  {Colmenarez}}, \bibinfo {author} {\bibfnamefont {V.}~\bibnamefont {Khemani}},
  \bibinfo {author} {\bibfnamefont {D.~J.}\ \bibnamefont {Luitz}},\ and\
  \bibinfo {author} {\bibfnamefont {D.~A.}\ \bibnamefont {Huse}},\ }\bibfield
  {title} {\bibinfo {title} {Avalanches and many-body resonances in many-body
  localized systems},\ }\href@noop {} {\bibfield  {journal} {\bibinfo
  {journal} {arXiv:2107.05642v1}\ } (\bibinfo {year} {2021}{\natexlab{b}})},\
  \Eprint {https://arxiv.org/abs/2107.05642} {arXiv:2107.05642
  [cond-mat.dis-nn]} \BibitemShut {NoStop}%
\bibitem [{\citenamefont {Igl\'oi}\ \emph {et~al.}(2012)\citenamefont
  {Igl\'oi}, \citenamefont {Szatm\'ari},\ and\ \citenamefont
  {Lin}}]{IgloiSzatmari}%
  \BibitemOpen
  \bibfield  {author} {\bibinfo {author} {\bibfnamefont {F.}~\bibnamefont
  {Igl\'oi}}, \bibinfo {author} {\bibfnamefont {Z.}~\bibnamefont
  {Szatm\'ari}},\ and\ \bibinfo {author} {\bibfnamefont {Y.-C.}\ \bibnamefont
  {Lin}},\ }\bibfield  {title} {\bibinfo {title} {Entanglement entropy dynamics
  of disordered quantum spin chains},\ }\href
  {https://doi.org/10.1103/PhysRevB.85.094417} {\bibfield  {journal} {\bibinfo
  {journal} {Phys. Rev. B}\ }\textbf {\bibinfo {volume} {85}},\ \bibinfo
  {pages} {094417} (\bibinfo {year} {2012})}\BibitemShut {NoStop}%
\bibitem [{\citenamefont {Vosk}\ and\ \citenamefont
  {Altman}(2013)}]{VoskAltman}%
  \BibitemOpen
  \bibfield  {author} {\bibinfo {author} {\bibfnamefont {R.}~\bibnamefont
  {Vosk}}\ and\ \bibinfo {author} {\bibfnamefont {E.}~\bibnamefont {Altman}},\
  }\bibfield  {title} {\bibinfo {title} {Many-body localization in one
  dimension as a dynamical renormalization group fixed point},\ }\href
  {https://doi.org/10.1103/PhysRevLett.110.067204} {\bibfield  {journal}
  {\bibinfo  {journal} {Phys. Rev. Lett.}\ }\textbf {\bibinfo {volume} {110}},\
  \bibinfo {pages} {067204} (\bibinfo {year} {2013})}\BibitemShut {NoStop}%
\bibitem [{\citenamefont {Pekker}\ \emph {et~al.}(2014)\citenamefont {Pekker},
  \citenamefont {Refael}, \citenamefont {Altman}, \citenamefont {Demler},\ and\
  \citenamefont {Oganesyan}}]{RefaelPekker}%
  \BibitemOpen
  \bibfield  {author} {\bibinfo {author} {\bibfnamefont {D.}~\bibnamefont
  {Pekker}}, \bibinfo {author} {\bibfnamefont {G.}~\bibnamefont {Refael}},
  \bibinfo {author} {\bibfnamefont {E.}~\bibnamefont {Altman}}, \bibinfo
  {author} {\bibfnamefont {E.}~\bibnamefont {Demler}},\ and\ \bibinfo {author}
  {\bibfnamefont {V.}~\bibnamefont {Oganesyan}},\ }\bibfield  {title} {\bibinfo
  {title} {Hilbert-glass transition: New universality of temperature-tuned
  many-body dynamical quantum criticality},\ }\href
  {https://doi.org/10.1103/PhysRevX.4.011052} {\bibfield  {journal} {\bibinfo
  {journal} {Phys. Rev. X}\ }\textbf {\bibinfo {volume} {4}},\ \bibinfo {pages}
  {011052} (\bibinfo {year} {2014})}\BibitemShut {NoStop}%
\bibitem [{\citenamefont {Zhao}\ \emph {et~al.}(2016)\citenamefont {Zhao},
  \citenamefont {Andraschko},\ and\ \citenamefont {Sirker}}]{Zhao2016}%
  \BibitemOpen
  \bibfield  {author} {\bibinfo {author} {\bibfnamefont {Y.}~\bibnamefont
  {Zhao}}, \bibinfo {author} {\bibfnamefont {F.}~\bibnamefont {Andraschko}},\
  and\ \bibinfo {author} {\bibfnamefont {J.}~\bibnamefont {Sirker}},\
  }\bibfield  {title} {\bibinfo {title} {Entanglement entropy of disordered
  quantum chains following a global quench},\ }\href
  {https://doi.org/10.1103/PhysRevB.93.205146} {\bibfield  {journal} {\bibinfo
  {journal} {Phys. Rev. B}\ }\textbf {\bibinfo {volume} {93}},\ \bibinfo
  {pages} {205146} (\bibinfo {year} {2016})}\BibitemShut {NoStop}%
\bibitem [{\citenamefont {Luitz}\ and\ \citenamefont
  {Lev}(2020)}]{BarLevLuitz}%
  \BibitemOpen
  \bibfield  {author} {\bibinfo {author} {\bibfnamefont {D.~J.}\ \bibnamefont
  {Luitz}}\ and\ \bibinfo {author} {\bibfnamefont {Y.~B.}\ \bibnamefont
  {Lev}},\ }\bibfield  {title} {\bibinfo {title} {Absence of slow particle
  transport in the many-body localized phase},\ }\href
  {https://doi.org/10.1103/PhysRevB.102.100202} {\bibfield  {journal} {\bibinfo
   {journal} {Phys. Rev. B}\ }\textbf {\bibinfo {volume} {102}},\ \bibinfo
  {pages} {100202} (\bibinfo {year} {2020})}\BibitemShut {NoStop}%
\bibitem [{\citenamefont {Abrahams}(2010)}]{50yearsAnderson}%
  \BibitemOpen
  \bibinfo {editor} {\bibfnamefont {E.}~\bibnamefont {Abrahams}},\ ed.,\ \href
  {https://doi.org/10.1142/7663} {\emph {\bibinfo {title} {50 Years of Anderson
  Localization}}}\ (\bibinfo  {publisher} {World Scientific},\ \bibinfo
  {address} {Singapore},\ \bibinfo {year} {2010})\BibitemShut {NoStop}%
\bibitem [{\citenamefont {Mossel}\ and\ \citenamefont
  {Caux}(2010)}]{Mossel_2010}%
  \BibitemOpen
  \bibfield  {author} {\bibinfo {author} {\bibfnamefont {J.}~\bibnamefont
  {Mossel}}\ and\ \bibinfo {author} {\bibfnamefont {J.-S.}\ \bibnamefont
  {Caux}},\ }\bibfield  {title} {\bibinfo {title} {Relaxation dynamics in the
  {gappedXXZspin}-1/2 chain},\ }\href
  {https://doi.org/10.1088/1367-2630/12/5/055028} {\bibfield  {journal}
  {\bibinfo  {journal} {New Journal of Physics}\ }\textbf {\bibinfo {volume}
  {12}},\ \bibinfo {pages} {055028} (\bibinfo {year} {2010})}\BibitemShut
  {NoStop}%
\end{thebibliography}
\end{document}